\documentstyle[aps,12pt,tighten]{revtex}
\begin{document}
\draft
\input{epsf}

\title{Search for Second Neutral Pion}

\author{W. A. Perkins}

\address{Perkins Advanced Computing Systems,\\ 12303 Hidden Meadows
Circle, Auburn, CA 95603, USA\\E-mail: wperkins@aub.com} 

\maketitle

\begin{abstract}
There is evidence of a second neutral pion from: (1) the anomalous
branching ratios in the reactions $\overline{p} \, p \rightarrow \pi \pi$ and 
$\overline{p}\, d \rightarrow \pi \pi N $, and (2) the 1960's results of 
Tsai-Ch\"{u} {\it et al.} for antinucleon annihilation stars in emulsions. 
The anomaly of (1) is eliminated if the two neutral pions in the reactions $\overline{p} \, p \rightarrow \pi^0 \, \pi^0$ and
$\overline{p}\, d \rightarrow \pi^0\, \pi^0\, n$ are not identical.
Tsai-Ch\"{u} {\it et al.} observed a second neutral pion that ``decays more rapidly into electron pairs with larger opening angles and more frequently into double pairs.'' One antineutron annihilation event produced three neutral particles (each with a mass of $135 \pm 14$ MeV), and each decayed into four electrons with much wider opening angles than those of the internal conversion electrons seen in $\pi^0$ decays. The larger opening angles and much more frequent double pair production could be caused by neutral pions with a lifetime so short that they sometimes decay into photon pairs before they can leave the annihilation nucleus (e.g., Ag) of the emulsion. We discuss several methods of searching for this second neutral pion.
\end{abstract}




\section{Introduction}
\label{sec.intro}

There is significant experimental evidence that capture and annihilation reactions occur predominately from atomic S states in liquid hydrogen.
The anomalously large fraction of antiproton annihilations proceeding from P states in two sets of reactions ($\overline{p}\, p \rightarrow \pi \pi $ and $\overline{p}\, d \rightarrow \pi \pi N$) is not measured directly, but is inferred using the theoretical argument that $\overline{p}\, p \rightarrow \pi^0 \pi^0 $ cannot occur from a $\overline{p}\, p$ atomic S state. However, if the two $\pi^0$'s are not identical, $\overline{p}\, p \rightarrow \pi^0_L \pi^0_S $ and $\overline{p}\, d \rightarrow \pi^0_L \pi^0_S n$  can occur from an atomic S states, and there is no anomaly. We denote the usual $\pi^0$ that decays in $\sim 10^{-16}$ s by $\pi^0_L$ and a second neutral pion that decays with a much shorter lifetime by $\pi^0_S$. 

The rules governing the $\overline{p}\, p \rightarrow \pi \pi$ reactions are discussed in Sec.~\ref{sec.allowed}, while the experimental evidence of the anomaly is given in Sec.~\ref{sec.expr}. In Sec.~\ref{sec.tsaichu} we present a slightly different interpretation of Tsai-Ch\"{u} {\it et al.} results. Their main point, that a second neutral pion with a very short lifetime exists, is unchanged. Whereas they assumed that the observed electrons came directly from the decay of this second neutral pion, $\pi^0_S$, we suggest that the observed electrons come from pair production inside the annihilation nucleus. Our interpretation has the advantage of explaining why such electron pairs and double pairs are not seen in $\overline{p}$ annihilation in hydrogen and deuterium. (These electron pairs are not Dalitz pairs as discussed in Sec.~\ref{sec.tsaichu}.)

Experimental tests that can show the existence of two distinct $\pi^0$ are discussed in Sec.~\ref{sec.tests} as well as another test (based on the results of Tsai-Ch\"{u} {\it et al.}) which can prove the existence and determine the lifetime of this second neutral pion.

\section{Allowed Antiproton-proton Reactions}
\label{sec.allowed}

The conservation of angular momentum, parity, and charge parity determine the allowed reactions in which protonium annihilates into two pions.
The eigenvalues of parity and charge parity of a 
fermion-antifermion pair are given
by~\cite{roman},

\begin{eqnarray}
\omega_{Parity} = (-1)^{X + 1}, \nonumber \\ 
\omega_{Charge \; parity} = (-1)^{X + s}, 
\label{eqnpbarp}
\end{eqnarray}
where $X $ is the relative orbital angular momentum of the 
two particles and $s$ is the spin of 
the fermion-antifermion system. The eigenvalues of parity and 
charge parity of a 
two pion system are given by~\cite{roman},

\begin{eqnarray}
\omega_{Parity} = (-1)^Y, \nonumber \\ 
\omega_{Charge \; parity} = (-1)^{Y}, 
\label{eqnpipi}
\end{eqnarray}
where $Y$ is the relative orbital angular momentum 
of the two pions. 

\begin{table}
\caption{ Initial state of $ \overline{p}\, p $  Atom }
\begin{tabular}{cccc}
\hline
State&$J$&Parity&Charge parity \\
\hline
$^1S_0$&0&-1&+1 \\
$^3S_1$&1&-1&-1 \\
$^3P_0$&0&+1&+1 \\
$^1P_1$&1&+1&-1 \\
$^3P_1$&1&+1&+1 \\
$^3P_2$&2&+1&+1 \\
\hline
\end{tabular}
\label{t1}
\end{table}

\begin{table}
\caption{ Final state of $  \pi^0 \pi^0 $ System }
\begin{tabular}{ccccc}
\hline
State&$J$&Parity&Charge parity&Comment \\
\hline
$S_0$&0&+1&+1& Y=0 \\
$D_2$&2&+1&+1& Y=2 \\
\hline
\end{tabular}
\label{t2}
\end{table}

\begin{table}
\caption{Final state of $  \pi^+ \pi^-$ System }
\begin{tabular}{ccccc}
\hline
State&$J$&Parity&Charge parity&Comment \\
\hline
$S_0$&0&+1&+1& Y=0 \\
$P_1$&1&-1&-1&Y=1  \\
$D_2$&2&+1&+1& Y=2 \\
\hline
\end{tabular}
\label{t3}
\end{table}
For the $ \pi^0 \pi^0 $ system there are further constraints. 
Because of Bose statistics, the state of two identical pions must 
be symmetric under interchange. Thus $Y$ must be even and 
both parity 
and charge parity must be $+1$ for the $ \pi^0 \pi^0 $ system. 
From these considerations
we obtain Tables~\ref{t1}$-$\ref{t3} for initial and final states.

Using conservation of parity, charge parity, and total angular 
momentum in matching the initial and final states, we determine the 
allowed reactions:

\begin{eqnarray}
\overline{p}\, p \: (^3S_1) \rightarrow \pi^+ \pi^- (P_1), \nonumber \\
\overline{p}\, p \: (^3P_0) \rightarrow \pi^0 \pi^0 (S_0), \nonumber 
\;\;\;\;\;\;
\overline{p}\, p \: (^3P_0) \rightarrow \pi^+ \pi^- (S_0), \nonumber \\
\overline{p}\, p \: (^3P_2) \rightarrow \pi^0 \pi^0 (D_2), \;\;\;\;\;\;
\overline{p}\, p \: (^3P_2) \rightarrow \pi^+ \pi^- (D_2). 
\label{eqnallow} 
\end{eqnarray}
Thus we see that the reaction $\overline{p}\, p \rightarrow  
\pi^0\pi^0 $ cannot occur from an atomic S state of the 
$\overline{p}\, p $ system if the two $\pi^0$'s are identical. In order to match the initial $^3S_1$ state of $\overline{p}\, p$, the
$\pi^0 \pi^0$ system needs a $P_1$ state with parity $= -1$ and 
charge parity $= -1$ 
as shown on the second line of Table~\ref{t3} for the 
$\pi^+ \pi^-$ system. 
The big difference in those eigenvalues between the
$\pi^+ \pi^-$ system and the $\pi^0 \pi^0$ system 
is caused by the two $\pi^0$'s being identical while the
$\pi^+$ and $\pi^-$ are not.

If the $\pi^0$'s are different, there is no requirement 
that the state of two $\pi^0$'s must be symmetric under interchange.  
Therefore, $Y$ can be odd for some $\pi^0_L \pi^0_S$ states
just as for the $\pi^+ \pi^-$ system, leading to a $P_1$ state with
parity of $-1$. The charge parity also
depends upon the relative angular momentum of
the two particles. Thus, the charge parity becomes
the same as in the $\pi^+ \pi^-$ case, given by 
Eq.~(\ref{eqnpipi}), and the reaction,
\begin{equation}
\overline{p}\, p \: (^3S_1) \rightarrow \pi^0_L \pi^0_S (P_1) 
\label{eqnpi0new} 
\end{equation}
is allowed. 

\section{Need for Two Neutral Pions to Explain Anomalous Branching Ratios}
\label{sec.expr}

We will briefly review what happens when an antiproton or some 
other negatively charged particle is slowed down in liquid  $H_2$ 
and is captured in a Bohr orbit by a proton. This is 
illustrated in Fig.~\ref{f1}. 
Typically, the incoming negatively charged 
particle is initially captured in an orbit with principle 
quantum number 
$ n \approx 30 $ and with high orbital angular momentum, $X$. 
Collisional deexcitations and radiative transitions transform 
the atom to 
lower $n$ and  $X$ values. The electrically neutral atom
can then penetrate 
neighboring atoms and experience the electric field of the protons.
This causes Stark effect transitions between 
the degenerate orbital angular momentum states. The rates for 
radiative transition and nuclear absorption (or annihilation)
from P states are small in 
comparison with the rate that the Stark effect 
populates the S state.
Since S-state absorption (or annihilation) 
can happen from high $n$ values,
the atom is unlikely to deexcite to low $n$ values for which P state
nuclear absorption (or annihilation) is more important. 


\epsfxsize=5.0in \epsfbox{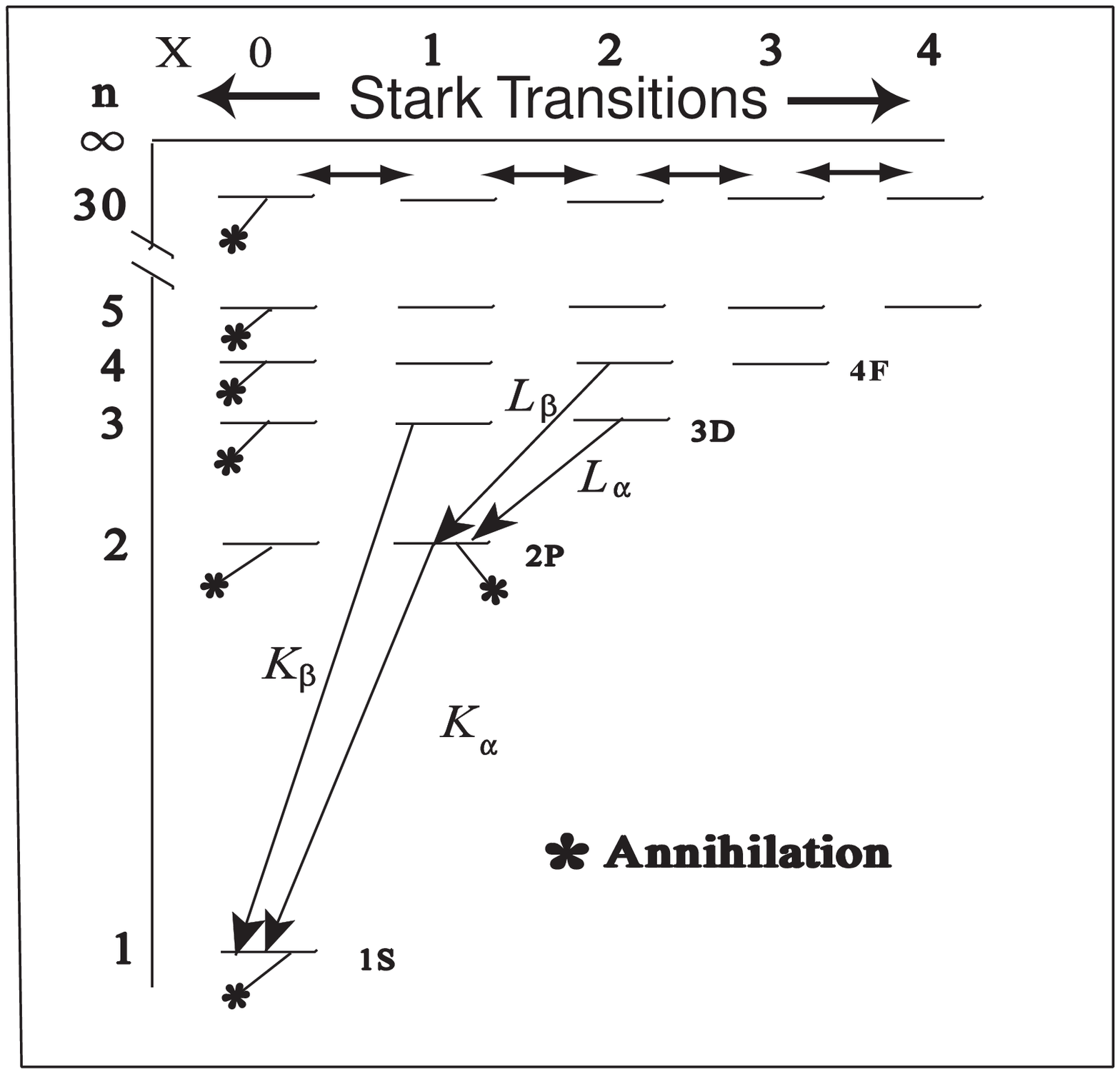}

\begin{figure}
\caption{Levels of atomic orbital states for a negatively charged 
particle and a proton with 
principle quantum number $n$ and orbital angular momentum $X$. 
It shows the effect of Stark transitions on different $X$ states, 
radiative deexcitations, and levels
from which nuclear absorption or annihilation are likely.}
\label{f1}
\end{figure}

Thus, according to theory~\cite{day,leon-bethe} absorption will 
occur predominantly from S states for $\pi^-$ and $K^-$. In 1960, 
Desai~\cite{desai} concluded, 
``Rough calculations indicate that for protonium 
also the capture will take place predominantly from S states.''

There is also strong experimental evidence that S-state capture dominates 
in liquid $H_2$. The reactions $ \pi^- p $~\cite{fields,bierman},  
$ K^- p $~\cite{cresti,knop}, and 
$ \Sigma^- p $~\cite{burnstein} have been studied. Since these 
negatively-charged particles decay, one can determine the nuclear 
absorption time by observing the fraction
which decay. The cascade times are about
two orders of magnitude shorter than would be required
for radiative deexcitation. 
Because the antiproton does not decay, such a measurement is not 
possible. Since  
the short cascade times for $\pi^-, K^-,$ and $\Sigma^-$ cannot be 
explained without recourse to the Stark effect, the Stark effect 
must also play a role in the $\overline{p}\, p$ case.

There is some direct evidence of S-state domination in
$\overline{p}\, p $ reactions. It has been determined~\cite{bizzarri1} 
that $\overline{p}\, p \rightarrow K K < 6\%$ from P states with a 95\%
confidence level. 
From the $\rho $ decay angular distribution, 
it has been determined~\cite{foster,bizzarri2} that  
$\overline{p}\, p \rightarrow  \pi^+ \pi^- \pi^0 < 5 \% $
from P states. Thus the experimental evidence strongly supports 
S-state domination for $\overline{p}\, p $ reactions. 

We will first look at the experimental results for 
$\overline{p} \,p \rightarrow  \pi \pi $. 
In liquid hydrogen the branching ratio for 
$\overline{p}\, p \rightarrow  \pi^+\pi^- $ 
is $ (32 \pm 1) \times 10^{-4}$~\cite{crystal2,crystal3}, 
while measurements of the branching ratio for 
$\overline{p}\, p \rightarrow  \pi^0\pi^0 $ are given in 
Table~\ref{t4}. 

The experimental results obtained
with the Crystal Barrel detector 
are likely to be the most accurate. 
As they noted in their paper~\cite{crystal1}, 
``Owning to our large detection efficiency and small background 
our result is least likely influenced by undetected systematic 
errors. The reliability of the result is strengthened by the 
internal consistency of a large set of two-body branching ratios 
measured with the Crystal Barrel detector and their 
agreement with previous determinations, especially with bubble 
chamber data.''

Assuming that the two $\pi^0$'s in the reaction 
$\overline{p}\, p \rightarrow \pi^0 \pi^0$ are identical, one can calculate the fraction of annihilations proceeding from P states, 
$P_{LH}(\pi \pi)$, for $ \overline{p}\, p \rightarrow \pi \pi $ as follows,

\begin{equation}
P_{LH}(\pi \pi)
= {BR( \overline{p} \,p \rightarrow \pi^+ \pi^-)_P
+ BR(\overline{p}\, p \rightarrow \pi^0 \pi^0)_P  \over
BR( \overline{p}\, p \rightarrow \pi^+ \pi^-)_{S \& P}
+ BR(\overline{p}\, p \rightarrow \pi^0 \pi^0)_P }.
\label{eqnfrac}
\end{equation}
Assuming charge independence,
\begin{equation}
BR( \overline{p}\, p \rightarrow \pi^+ \pi^-)_P
= 2 \times BR(\overline{p}\, p \rightarrow \pi^0 \pi^0)_P,
\label{eqnchind}  
\end{equation}
we obtain the \% proceeding from P states that is given in Column 2 of Table~\ref{t4}. The result of 48\% proceeding from P states, shown in Table~\ref{t4} (Crystal Barrel collaboration), is anomalously high. 

\begin{table}
\caption{ Branching Ratio for
 $ \overline{p}\, p \rightarrow \pi^0 \pi^0 $  }
\begin{tabular}{cccc}
\hline
Measured value& \% from P states&Year&Reference \\
\hline
$(4.8 \pm 1.0) \times 10^{-4}$&$39\%$&1971&Devons {\it et al.}
~\cite{devons} \\
$(1.4 \pm 0.3)\times 10^{-4}$&$13\%$&1979&Bassompierre {\it et al.}
~\cite{bassom}\\
$(6 \pm 4) \times 10^{-4}$&$47\%$&1983&Backenstoss {\it et al.}
~\cite{backen} \\
$(2.06 \pm 0.14) \times 10^{-4}$&$18\%$&1987&Adiels {\it et al.}
~\cite{adiels} \\
$(2.5 \pm 0.3) \times 10^{-4}$&$22\%$&1988&Chiba {\it et al.}
~\cite{chiba}\\
$(6.93 \pm 0.43) \times 10^{-4}$&$53\%$&1992&Crystal Barrel
~\cite{crystal1} \\
$(2.8 \pm 0.4) \times 10^{-4}$&$24\%$&1998&Obelix
~\cite{obelix} \\
$(6.14 \pm 0.40) \times 10^{-4}$&$48\%$&2001&Crystal Barrel
~\cite{crystal3} \\
\hline
\end{tabular}
\label{t4}
\end{table}

We now consider antiproton annihilation in deuterium. From studying the reactions $\overline{p} \, d \rightarrow  \pi^-\pi^+ n $
and  $\overline{p} \, d \rightarrow  \pi^-\pi^0 p $
in a liquid deuterium bubble chamber, Gray {\it et al.}~\cite{gray} reported 
that $(75 \pm 8)\%$ of the annihilations come from P states. The quantity measured is,
\begin{equation}
r = {BR( \overline{p} \, d \rightarrow \pi^- \pi^0 p) \over
BR( \overline{p} \, d \rightarrow \pi^+ \pi^- n) }.
\label{eqndeutratio}
\end{equation}
The percentage proceeding from P states was then calculated using charge independence,
\begin{equation}
BR( \overline{p} \, d \rightarrow \pi^+ \pi^- n)
= {1 \over 2} BR( \overline{p}\, d \rightarrow \pi^- \pi^0 p)
+ 2 \times BR(\overline{p}\, d \rightarrow \pi^0 \pi^0 n),
\label{eqnchind2}  
\end{equation}
and 
\begin{equation}
P_{LD}(\pi \pi) 
=  {3BR(\overline{p}\, d \rightarrow \pi^0 \pi^0 n)  \over
BR( \overline{p}\, d \rightarrow \pi^+ \pi^- n)
+ BR(\overline{p} \, d \rightarrow \pi^0 \pi^0 n) },
\label{eqndeutfrac}
\end{equation}
resulting in,
\begin{equation}
P_{LD}(\pi \pi) =  {3(2 - r) \over {6 - r}}.
\label{eqndeutrcalc}
\end{equation}
Equation (\ref{eqndeutfrac}) is based on the theoretical argument that $\overline{p} \, d \rightarrow  \pi^0\pi^0 n$ cannot occur from an atomic S state. This argument is identical to that discussed in Sec.~\ref{sec.allowed} for the reaction 
$\overline{p}\, p \rightarrow \pi^0 \pi^0 $.

The results of Gray {\it et al.}~\cite{gray} and two other groups are given in 
Table~\ref{t5}.
The experimental results of Bridges {\it et al.}~\cite{bridges} 
using a magnetic spectrometer are in close agreement with those of
Gray {\it et al.}~\cite{gray}, but the high statistics experiment of
Angelopoulos {\it et al.}~\cite{angel} using a magnetic spectrometer is consistent with a P-state fraction of 0\%. Reifenr\"{o}ther and Klempt~\cite{reinfen} have noted that this low value~\cite{angel} could have been caused by the tight cut on the colinearity of the $\pi^+\pi^-$ pair that was used. Too tight a cut can result in  $\pi^+\pi^-$ pairs being lost. Gray {\it et al.}~\cite{gray} used a cut of $16$ degrees, Bridges {\it et al.}~\cite{bridges} used $10$ degrees, and 
Angelopoulos {\it et al.}~\cite{angel} used $5$ degrees.

\begin{table}
\caption{ Ratio 
 ${BR( \overline{p} \,d \rightarrow \pi^- \pi^0 p)} / 
{BR( \overline{p}\, d \rightarrow \pi^+ \pi^- n) }$ }
\begin{tabular}{ccccc}
\hline
Measured value& Method& \% from P states&Year&Reference \\
\hline
$(0.68 \pm 0.07)$ &deuterium&$75\%$&1973&Gray {\it et al.}
~\cite{gray} \\
&bubble&&&\\
&chamber&&&\\
$(0.70 \pm 0.05)$&magnetic&$74\%$&1986&Bridges {\it et al.}
~\cite{bridges} \\
$(0.55 \pm 0.05)$&spectrometer&$80\%$&1986&Bridges {\it et al.}
~\cite{bridges} \\
$(2.07 \pm 0.05)$&magnetic&$0\%$&1988&Angelopoulos {\it et al.}
~\cite{angel} \\
&spectrometer&&&\\
\hline
\end{tabular}
\label{t5}
\end{table}


\epsfxsize=5.0in \epsfbox{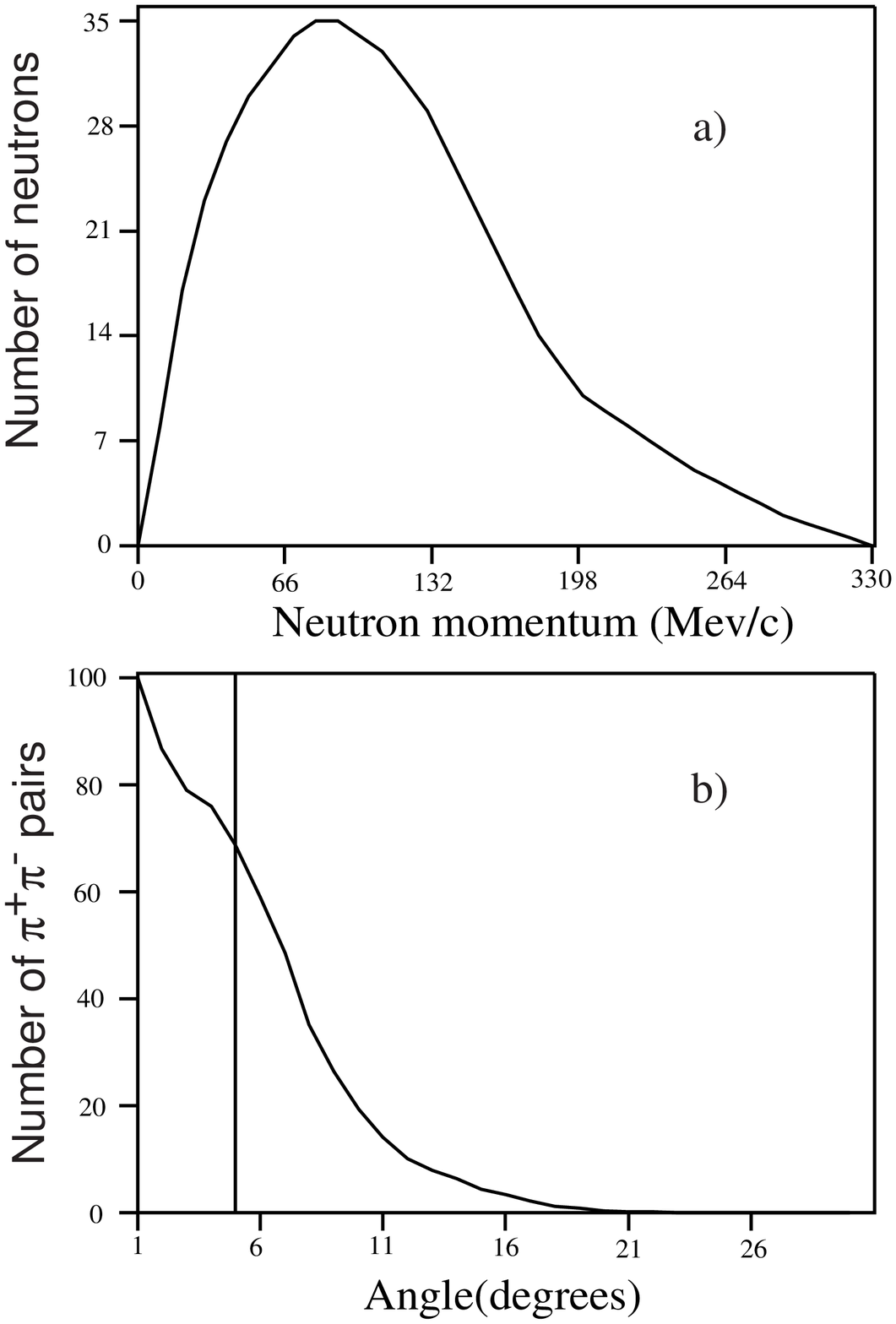}

\begin{figure}
\caption{
(a) Neutron momentum distribution in reaction 
$\overline{p}\, d \rightarrow \pi^+ \pi^- n$ 
at rest, derived from result of Gray et al. [22]. 
(b) Monte Carlo calculation of number of $\pi^+ \pi^-$ pairs versus 
their angular deviation from colinearity in reaction $\overline{p} \,d \rightarrow \pi^+ \pi^- n$. Vertical line is at 5 degree cutoff.} 
\label{f2}
\end{figure}

We did a Monte Carlo calculation of the expected deviation from colinearity of the $\pi^+\pi^-$ pairs for the reaction $\overline{p} \, d \rightarrow  \pi^+\pi^- n$. We used the neutron momentum distribution shown in Fig.~\ref{f2}a, that was derived from the results of Gray {\it et al.} ~\cite{gray}.
(See Fig. 1c of Ref.~\cite{gray} and Fig.~5 of Ref.~\cite{bridges}.) The calculated angular deviation from colinearity, plotted in Fig.~\ref{f2}b, shows that a significant fraction of the pairs extending beyond 5 degrees. Our correction factor for pairs beyond 5 degrees is 1.58, while that used in 
Ref.~\cite{angel} was 1.13. Using this new correction factor results in 
$r = 1.48$ and $ P_{LD}(\pi \pi) = 34\%$, which is in better agreement with the
first two experiments.

Reifenr\"{o}ther and Klempt~\cite{reinfen} have suggested a modification which includes the measured ratio (1.33) of $\overline{p}\, p$ to $\overline{p}\, n$ in $D_2$. This reduces the $75\%$ from P states to $55\%$. With considerations similar to those of Eqs.~(\ref{eqnfsubp1})$-$(\ref{eqnfsubp3}) shown below, they obtain an $f_P$ of $45\%$. 

Some efforts have been made to eliminate the discrepancy in $\overline{p}\, p \rightarrow \pi \pi$ by lowering the P-state fraction inferred from the measured branching ratios. Doser {\it et al.}~\cite{doser} suggested that the total P-state fraction might be lower than that given by Eq.~(\ref{eqnfrac}), and it was just that the branching ratio into two pions (rather than into other particles) is greater from P states than it is from S states. In support of that idea, Doser {\it et al.}~\cite{doser} measured the branching ratio for 
$\overline{p}\, p \rightarrow \pi^+ \pi^-$ from a pure P state (by detection of the reaction in coincidence with transition x-rays) to be $(4.81 \pm 0.49) \times 10^{-3}$. 

One can calculate the P-state annihilation fraction, $f_P$, using~\cite{reinfen,batty},
\begin{eqnarray}
BR(\overline{p}\, p \rightarrow \pi^+ \pi^-) = (1 - f_P)B_0 + f_P B_1, 
\label{eqnfsubp1}  \\
BR(\overline{p}\, p \rightarrow \pi^0 \pi^0) = {1 \over 2}f_P B_1, 
\label{eqnfsubp2} \\
BR(\overline{p}\, p \rightarrow \pi^+ \pi^-)_X = B_1.
\label{eqnfsubp3} 
\end{eqnarray}
where $B_0$ and $B_1$ are the branching-ratio coefficients (independent of density) from S and P states respectively.

Using the Crystal Barrel collaboration result~\cite{crystal3} and Doser {\it et al.} result~\cite{doser} for $BR( \overline{p}\, p \rightarrow \pi^+ \pi^-)_X$, one obtains $B_1 = 4.81 \times 10^{-3}$ from Eq.~(\ref{eqnfsubp3}) and $f_P = 26\%$ from (\ref{eqnfsubp2}).

In an attempt to further reduce this percentage, Batty~\cite{batty} discovered an interesting mechanism. He introduced enhancement of annihilations
from fine structure states over that expected from
a statistical population. He modified the earlier 
calculations of Reifenr\"{o}ther and Klempt~\cite{reinfen}
using the Borie and Leon model~\cite{borie} to incorporate enhancement factors.  With enhancement factors Eqs.~(\ref{eqnfsubp1})$-$(\ref{eqnfsubp3}) become,
\begin{eqnarray}
BR(\overline{p}\, p \rightarrow \pi^+ \pi^-) 
= (1 - f_P){3 \over 4} E(^3S_1) B(^3S_1) \nonumber \\  
+ f_P \left[ {1 \over 12} E(^3P_0) B(^3P_0)
+ {5 \over 12} E(^3P_2) B(^3P_2) \right], \\
\label{eqn313} 
BR(\overline{p}\, p \rightarrow \pi^0 \pi^0)  
= {1 \over 2} f_P \left[ {1 \over 12} E(^3P_0) B(^3P_0)
+ {5 \over 12} E(^3P_2) B(^3P_2) \right], \\
\label{eqn314} 
BR(\overline{p}\, p \rightarrow \pi^+ \pi^-)_X 
= {1 \over 12} B(^3P_0)
+ {5 \over 12} B(^3P_2).
\label{eqn315} 
\end{eqnarray}

The enhancement factors $E(^3S_1)$ and $E(^3P_2) \approx 1.0$ for all densities while the enhancement factor for the $^3 P_0$ state, $E({^3 P_0}) \approx 1.0$ at low density and increases to $2.076$ to $2.556$ (depending upon the model used in the calculation) at liquid $H_2$ density. In order to reduce $f_P$, Batty assumed $B({^3 P_0}) \gg B({^3 P_2})$, but his result, $f_P = 10$ to $12\%$, is still at least a factor of 2 too high. The reduction depends strongly on the choice of parameters. For example, if we take $B({^3 P_0}) = B({^3 P_2})$, $f_p$ would only be reduced to $20\%$. Thus, Batty's mechanism is not large enough to remove the discrepancy for the $\overline{p}\, p \rightarrow \pi \pi$ case. In addition, enhancement factors are not effective in reducing the discrepancy in the $\overline{p}\, d \rightarrow \pi \pi N$ case~\cite{batty2} as the predicted enhancement factors are approximately 1.

In summary, if we assume that the two $\pi^0$'s are identical, the branching ratios for the reactions $\overline{p} \,p \rightarrow \pi \pi $ 
and $\overline{p}\,d \rightarrow \pi \pi N$ indicate a fraction proceeding from P states that is a factor of 4 to 8 greater than that occurring in other reactions. However, if the two $\pi^0$'s are not identical, the reactions can occur from S states and the anomaly is eliminated. 

\section{New Interpretation of Tsai-Ch\"{u} {\it et al.} Results}
\label{sec.tsaichu}

In the 1960's Tsai-Ch\"{u} {\it et al.}~\cite{tsai2,tsai}  found evidence of a second neutral pion with some surprising properties. They placed stacks of K-5 emulsions in the antiproton beam of the Berkeley Bevatron and looked for multi-prong stars. They were surprised to see many electrons coming from some of the annihilation vertices. 

By charge exchange some of the antiprotons are converted to antineutrons. One star caused by an antineutron annihilation produced 12 electrons~\cite{tsai2}. The analysis showed that electrons (four each) came from three neutral particles, with masses of $136 \pm 14$, $135 \pm 14$, and $136 \pm 13$ MeV. The likelihood of three ordinary neutral pions decaying with double Dalitz pairs is less than $10^{-13}$.

From analysis of 15 antinucleon annihilation stars, Tsai-Ch\"{u} {\it et al.}~ \cite{tsai} reported the following properties for this second neutral pion: (1) It has a mass of the same order as the usual $\pi^0$, (2) It is emitted with the same energy as that of a charged pion, (3) It decays more often into electron pairs and into double pairs, (4) The electron pairs from this second $\pi^0$ have larger opening angles than those of Dalitz pairs, and (5) It has a very short lifetime (much shorter than the usual $\pi^0$) because the electrons are emitted directly from the origin of the stars.

Recent experiments with much improved low-energy antiproton beams annihilating in liquid hydrogen and deuterium have not shown emission of electron pairs with those characteristics. What was happening in those experiments of Tsai-Ch\"{u} {\it et al.}? A simple explanation is that this second neutral pion has a lifetime so short that it occasionally decays before it can leave the annihilation nucleus of the emulsion, e.g., an Ag nucleus. This is illustrated in Fig.~\ref{f3}.

The main decay mode of this second neutral pion must be $\pi^0_S \rightarrow 2 \gamma$ because this is the way each $\pi^0$ is detected~\cite{crystal3} in  
$\overline{p} \, p \rightarrow \pi^0_L \, \pi^0_S$ experiments. The probability for creation of electron pairs by $\pi^0$ photons is very high inside a nucleus. This explains the appearance of electrons in heavy-nuclei annihilations but not in hydrogen and deuterium annihilations.
The opening angles for pairs produced by high-energy photons on nuclei~\cite{holtwijk} are wider than those from Dalitz pairs and the distribution of angles is in reasonable agreement with the result given in Table III of Tsai-Ch\"{u} {\it et al.}~\cite{tsai}. 


\epsfxsize=5.0in \epsfbox{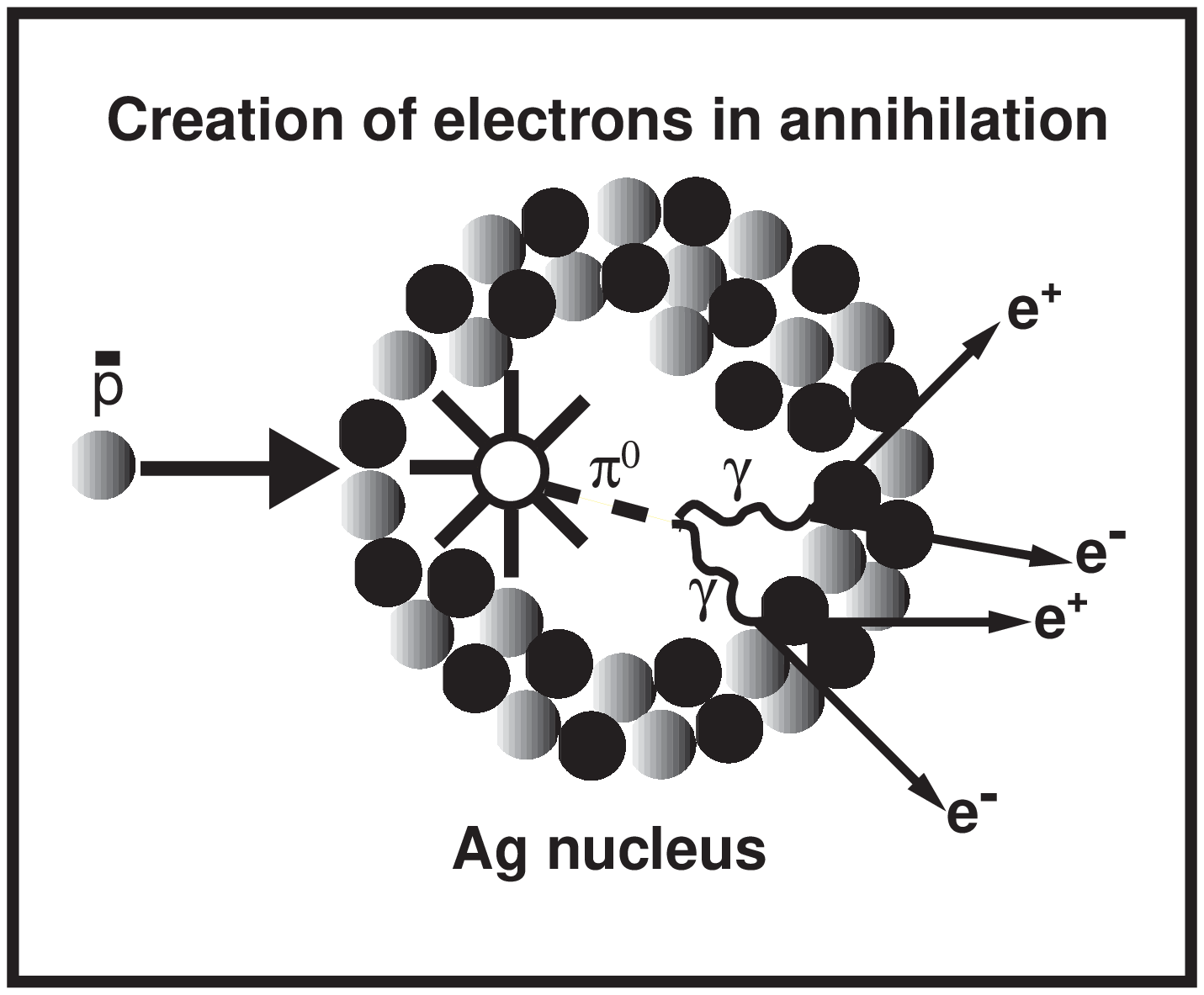}
  
\begin{figure}
\caption{Illustration of proposed method by which second neutral pion
appears to decay into four electrons. The $\pi^0_S$ decays inside the annihilation nucleus and its photons produce the observed electrons by pair production. Since $\pi^0_S$ must decay inside
the nucleus, this process requires a lifetime $\sim 10^{-21}$ s or less.}
\label{f3}
\end{figure}

From the results of Tsai-Ch\"{u} {\it et al.} it appears that the annihilations that produce these electrons occur in 1 to 10\% of the annihilation stars. Assuming one $\pi^0_S$ per star, one can estimate a lifetime between $10^{-21}$ s and $10^{-22}$ s. From the uncertainty principle, the corresponding width is between $0.7$ MeV and $7$ MeV.  The lifetime of $\pi^0_S$ cannot be much shorter than $10^{-22}$ s, or it would have been noticed in missing mass spectra for reactions such as $\overline{p} \, p \rightarrow \pi^+ \, \pi^- \, \pi^0_S$. A wide resonance at $m_{\pi^0}$ would look significantly different than the detected, narrow  $\pi^0$ peak. 

\section{Experimental Tests}
\label{sec.tests}

Although seven different groups have measured the branching ratio
for $\overline{p}\, p \rightarrow \pi^0 \pi^0 $, showing the importance
of this unexpectedly large branching ratio, no 
direct test has been performed to determine 
whether the reaction could be occurring from an
atomic S state. Such an experimental test can be made by setting 
up an initial $\overline{p}\, p $ atomic S state and looking for 
the $\pi^0 \pi^0$ final state. 
The method is illustrated in Fig.~\ref{f4}. 
As discussed earlier, in liquid $H_2$ the Stark effect causes 
transitions to S states at high n-values, where annihilation 
occurs more readily than deexcitation. One can 
decrease the effect of Stark transitions by using $H_2$ gas 
at STP, and thereby observe the deexcitation radiation.


\epsfxsize=5.0in \epsfbox{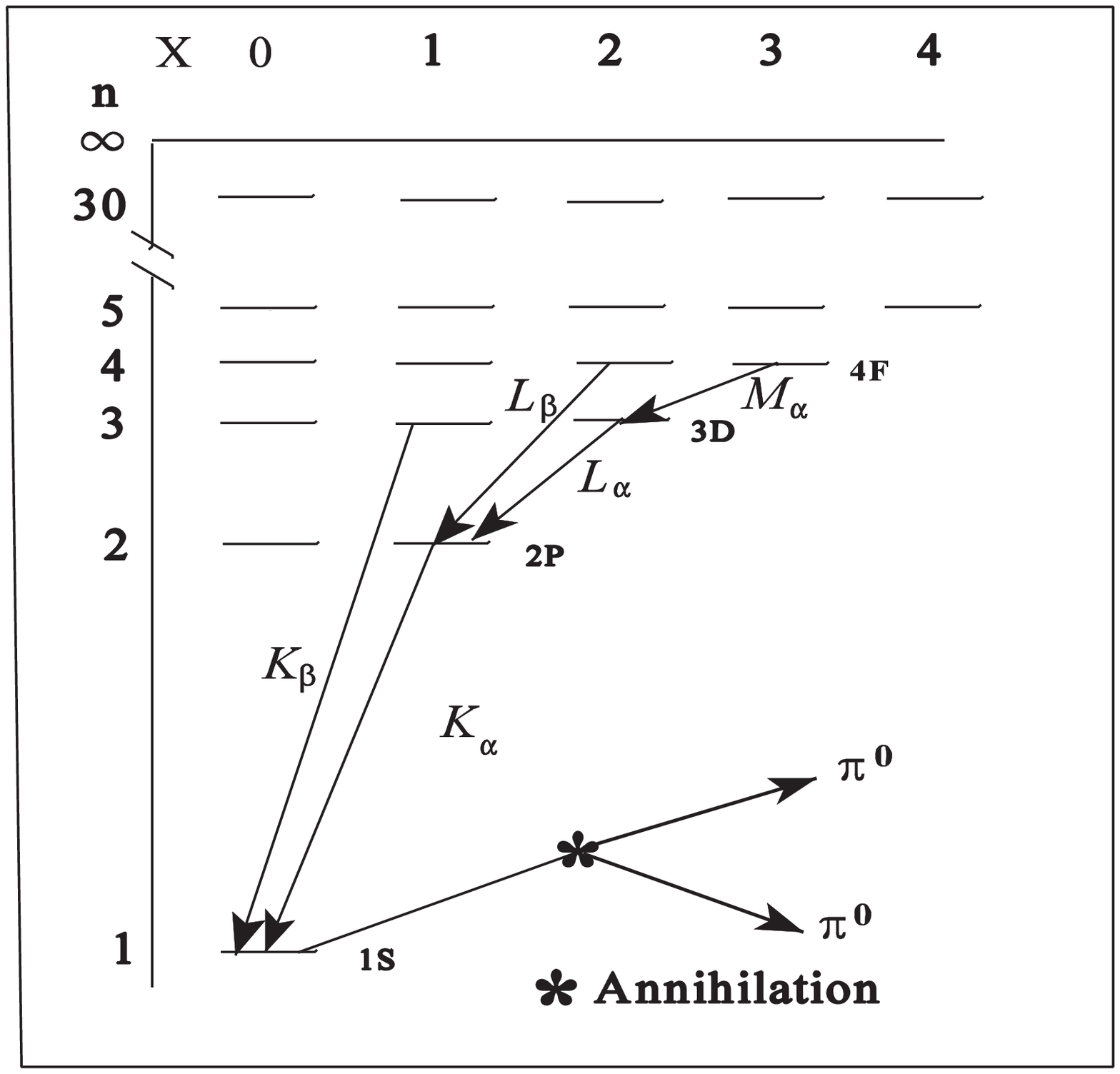}

\begin{figure}
\caption{Test for $\overline{p}\, p \rightarrow \pi^0_L \pi^0_S $ from S state. 
It shows radiative cascades to the 1S state which 
involve L and K X-rays followed by annihilation.}
\label{f4}
\end{figure}

The coincidence of L and K X-rays from protonium, shows that 
the atom is in the 1S state. The energy of the K X-rays is 
between 9.4 KeV ($K_{\alpha}$) and 12.5 KeV ($K_{\infty}$), 
while energy of the L X-rays is between 1.7 KeV and 3.1 KeV. 
The energy of M X-rays is between 0.5 KeV and 1.3 KeV. Thus, 
the X-rays from the different transitions tend to be separated. 

The Asterix Collaboration has detected $K_{\alpha}$ X-rays in 
coincidence with L X-rays~\cite{asterix2}. The experiment 
we are proposing 
is very similar, but requires the triple coincidence of L and K 
X-rays from protonium and the $\pi^0 \pi^0$ annihilation mode. 
The detection of such events can prove that the annihilation 
reaction is occurring from an atomic S state.

A referee has pointed out that some vector mesons 
can decay into $\pi^0_L \, \pi^0_S$ if the $^3S_1$ state of protonium can,
because certain vector mesons also have $J = 1$, 
parity $= -1$, and charge parity $= -1$. 
We considered the $\pi^0_L \pi^0_S$ decay mode of the $\rho(770)^0$,
$\omega(782)$, $\phi(1020)$, and $J/\psi(1S)$. 
This decay mode of the $\omega(782)$, $\phi(1020)$, and $J/\psi(1S)$ is forbidden by G-parity conservation, but it can proceed electromagnetically.
In estimating the branching ratios, we assumed that the 
$\pi^0_L \pi^0_S$ mode would occur at about the same rate as the $\pi^+\pi^-$ 
mode, but reduced by a factor of two. 

The $\rho(770)^0$ has $I = 1$, so the reaction $\rho(770)^0 \rightarrow \pi^+\pi^-$ is allowed by isospin conservation because 
the $\pi^+\pi^-$ system can have I $=$ 0, 1, or 2. 
The reaction $\rho(770)^0 \rightarrow \pi^0\pi^0$ 
is forbidden for the usual $\pi^0$ because the $\pi^0\pi^0$ system can only have I $=$ 0 or 2. Occurring electromagnetically, its branching ratio is reduced 
by the factor $\alpha^2$. Assuming that $\pi^0_S$ has isospin 1, then
the reaction $\rho(770)^0 \rightarrow \pi^0_L\pi^0_S$ is 
similarly suppressed. This assumption is crucial for otherwise the reaction would have been readily observed.
 
Our estimated branching ratios are,
\begin{eqnarray}
BR(\rho(770)^0 \rightarrow  \pi^0_L\pi^0_S) = 5 \times 10^{-5}, \nonumber \\
BR(\omega(782) \rightarrow  \pi^0_L\pi^0_S) = 1 \times 10^{-2}, \nonumber \\
BR(\phi(1020) \rightarrow \pi^0_L\pi^0_S) = 4 \times 10^{-5}, \nonumber \\
BR(J/\psi(1S) \rightarrow  \pi^0_L\pi^0_S) = 7 \times 10^{-5}.
\label{branching}
\end{eqnarray}
The only measured 
upper limit~\cite{achasov},
$BR(\phi(1020) \rightarrow  \pi^0\pi^0) 
< 4 \times 10^{-5}$, is just in the expected range. Search for the decay mode,
$\omega(782) \rightarrow \pi^0_L\pi^0_S$ with an expected branching ratio of $1 \times 10^{-2}$ seems very attractive. It is already known that $\omega(782)$ decays into undetermined neutrals with a branching ratio between  $1.8 \times 10^{-3}$ and
$1.4 \times 10^{-2}$.

The results of Tsai-Ch\"{u} {\it et al.}~\cite{tsai2,tsai} suggest another test that could allow a determination of the lifetime of this second $\pi^0$. By varying the mass number, A, of the target material in low-energy antiproton annihilation, one should be able to observe the increase in electrons (as A increases) from the decay of $\pi^0_S$'s inside the annihilation nucleus as indicated by Fig.~\ref{f5}. By measuring the absolute increase in magnitude and variation with A, one should be able to determine the lifetime of $\pi^0_S$.

\section{Conclusions}
\label{sec.disc}

This has been a phenomenology investigation. The anomalous results for the reactions $\overline{p}\, p \rightarrow \pi \pi $ 
and $\overline{p}\, d \rightarrow \pi \pi N$ were a strong indication of the existence of two distinct $\pi^0$'s. The results of Tsai-Ch\"{u} {\it et al.} 
gave direct evidence of a different kind of $\pi^0$. With our reinterpretation of their results, it is not surprising that this second $\pi^0$ has not been noticed in other reactions, because the major difference between it and the other $\pi^0$ is its much shorter lifetime.

We do not have a theoretical model for this second neutral pion with its internal quark structure and its relationship to the other pions. It is tempting to think that the two neutral pions, $\pi^0_L$ and $\pi^0_S$, are related to each other as $K^0_L$ and $K^0_S$, but there are difficulties in forming such a model. The present evidence for two distinct neutral pions comes from antinucleon-annihilation reactions, but we believe that most other reactions produce $\pi^0_L$ and $\pi^0_S$ in equal abundance as occurs for $K^0_L$ and $K^0_S$.


\epsfxsize=5.0in \epsfbox{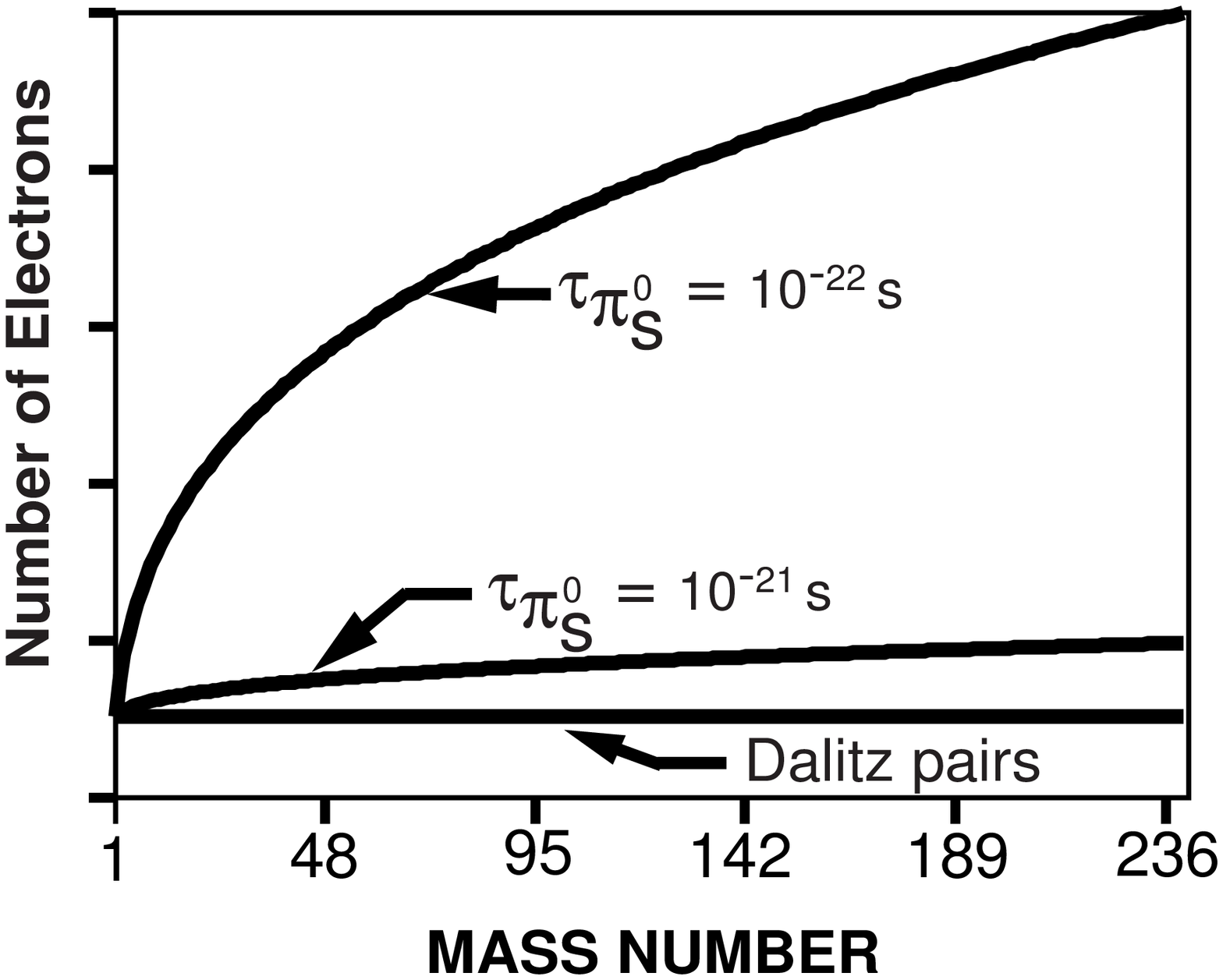}

\begin{figure}
\caption{Expected variation in number of electrons produced by
antiproton annihilation versus mass number for two different
lifetimes of second neutral pion.} 
\label{f5}
\end{figure}

Since the main difference between the two neutral pions is that one has a lifetime $\sim 10^{-16}$~s while the other has a lifetime from $10^{-21}$~s to $10^{-22}$~s, one might think that this short-lived $\pi^0$ would have been detected in the $\pi^0$ lifetime measurements~\cite{caso}. However, because the usual $\pi^0$ has such a short lifetime, it is difficult to separate it from one with a shorter lifetime. In addition, since the experimenters were not looking for a prompt decaying $\pi^0_S$, they often worked to eliminate prompt signals as unwanted background.

For example, the experiment of Atherton {\it et al.}~\cite{Atherton} was designed so the measurement of the mean decay length would not be affected by prompt decays such as $\eta \rightarrow \gamma \gamma$. In their ratio 
R = [Y(250) - Y(45)] /  [Y(250) - Y(0)] the positrons from prompt decays, which are not dependent on foil separation, are cancelled out. Shwe {\it et al.}~\cite{shwe} apparently ignored prompt decays to eliminated confusing background. As they noted, ``Among the events missed or unmeasured were...Events with very small gaps which lie within the `circle of confusion' around the star center.'' Stamer {\it et al.} ~\cite{stamer} worked with the
$K^+ \rightarrow \pi^+ \pi^0$ decay at rest, and their histogram of number of decays versus gap shows a very large peak in the 0 to 0.5 micron bin that could contain half prompt decays by a short-lived $\pi^0_S$. One of the most accurate methods of measuring the $\pi^0$ lifetime is based on the Primakoff effect, involving coherent photoproduction of $\pi^0$'s in the coulomb field of nuclei. Unlike the other techniques, this method is heavy on theory. Although the results do not indicate a short-lived $\pi^0$, this method may not be appropriate for a second $\pi^0$ that decays by some mechanism in the $10^{-21}$ s to $10^{-22}$ s range. 

Searching for this second neutral pion is of utmost importance. We have suggested some tests in Sec.~\ref{sec.tests} which can prove the existence of two distinct $\pi^0$'s and some tests to confirm the existence of a $\pi^0_S$ with a very short lifetime.

\acknowledgments

Helpful discussions with Prof. J. E. Kiskis are 
gratefully acknowledged.


\begin{references}

\bibitem{roman} 
P.~Roman, {\it Theory of Elementary Particles} (North-Holland, Amsterdam, 
1960) pp.~287--305.

\bibitem{day}
T.~B.~Day, G.~A.~Snow, and J.~Sucher, Phys. Rev. Lett. {\bf 3}, 61 (1959).

\bibitem{leon-bethe}
M.~Leon and H.~A.~Bethe, Phys. Rev. {\bf 127}, 636 (1962).

\bibitem{desai} 
B.~R.~Desai, Phys. Rev. {\bf 119}, 1385 (1960).

\bibitem{fields}
T.~H.~Fields, G.~B.~Yodh, M.~Derrick, and J.~G.~Fetkovich, 
Phys. Rev. Lett. {\bf 5}, 69 (1960).

\bibitem{bierman}
E.~Bierman, S.~Taylor, E.~L.~Koller, P.~Stamer, and T.~Huetter, Phys. Lett. {\bf 4}, 351 (1963).

\bibitem{cresti}
M.~Cresti, S.~Limentani, A.~Loria, L.~Peruzzo, and R.~Santangelo, Phys. Rev. Lett. {\bf 14}, 847 (1965).

\bibitem{knop}
R.~Knop, R.~A.~Burnstein, and G.~A.~Snow, 
Phys. Rev. Lett. {\bf 14}, 767 (1965).

\bibitem{burnstein}
R.~A.~Burnstein, G.~A.~Snow, and H.~Whiteside, Phys. Rev. Lett.
{\bf 15}, 639 (1965).

\bibitem{bizzarri1} 
R.~Bizzarri, G.~Ciapetti, U.~Dore, M.~Gaspero, I.~Laakso, F.~Marzano,
And G.~C.~Moneti, Nucl. Phys. {\bf B 69}, 307 (1974).

\bibitem{foster}
M.~Foster, P.~Gavillet, G.~Labrosse, L.~Montanet, R.~A.~Salmeron,
P.~Villemoes, C.~Ghesquiere, and E.~Lillestol, Nucl. Phys. {\bf B 6}, 107 (1968).

\bibitem{bizzarri2}
R.~Bizzarri, in {\it Symp. on Nucleon-Antinucleon Annihilations, ed.
L.~Montanet, Chexbres, Switzerland} (CERN Report TC/72-10, 1972)
p.~161.

\bibitem{crystal2} 
Crystal Barrel Collaboration, C.~Amsler {\it et al.}, 
Z. Phys. {\bf C 58}, 175 (1993).

\bibitem{crystal3} 
Crystal Barrel Collaboration, A.~Abele {\it et al.}, 
Nucl. Phys. {\bf A 679}, 563 (2001).

\bibitem{devons}
S.~Devons, T.~Kozlowski, P.~Nemethy, S.~Shapiro, N.~Horwitz, T.~Kalogeropoulos,
J.~Skelly, R.~Smith, and H.~Uto, Phys. Rev. Lett. {\bf 27}, 1614 (1971).

\bibitem{bassom} 
G.~Bassompierre {\it et al.}, 
{\it Proc. 4th European Antiproton Conf., Vol. I, Strasbourg}, ed.
A.~Freidman, (CNRS, Paris, 1979) p.~139.

\bibitem{backen} 
G.~Backenstoss {\it et al.}, Nucl. Phys. {\bf B 228}, 424 (1983).

\bibitem{adiels}
L.~Adiels {\it et al.}, Z. Phys. {\bf C 35}, 15 (1987).

\bibitem{chiba}
M.~Chiba {\it et al.}, Phys. Rev. {\bf D 38}, 2021 (1988).

\bibitem{crystal1} 
Crystal Barrel Collaboration, C.~Amsler {\it et al.}, 
Phys. Lett. {\bf B 297}, 214 (1992).

\bibitem{obelix} 
Obelix Collaboration, A.~Zoccoli {\it et al.}, 
{\it Proc. HADRON 97}, eds.  S-U.~Chung and H.~J.~Willutzki, AIP
conf. proc. 432 (1998) p.~347.

\bibitem{gray}
L.~Gray, T.~Papadopoulou, E.~Simopoulou, A.~Vayaki, T.~Kalogeropoulos, and 
J.~Roy, Phys. Rev. Lett. {\bf 30}, 1091 (1973).

\bibitem{bridges}
D.~Bridges {\it et al.}, Phys. Lett. {\bf B 180}, 313 (1986).

\bibitem{angel} 
A.~Angelopoulos {\it et al.}, Phys. Lett. {\bf B 212}, 129 (1988).

\bibitem{reinfen}
G.~Reifenr\"{o}ther and E.~Klempt, Phys. Lett. {\bf B 245}, 129 (1990).

\bibitem{doser}
ASTERIX Collaboration, M.~Doser {\it et al.},
 Nucl. Phys. {\bf A 486}, 493 (1988).

\bibitem{batty} 
C.~J.~Batty, Nucl. Phys. {\bf A 601}, 425 (1996).

\bibitem{borie}
E.~Borie and M.~Leon, Phys. Rev. {\bf A 21}, 1460 (1980).

\bibitem{batty2} 
C.~J.~Batty, Nucl. Phys. {\bf A 655}, 203c (1999).

\bibitem{tsai2}
Tsai-Ch\"{u} and C.~Simonin-Haudecoeur, Nuovo Cimento {\bf 20}, 1102 (1961).

\bibitem{tsai}
Tsai-Ch\"{u}, C.~Simonin-Haudecoeur, D.~Schune-Boisble, and B.~Brami-Depaux, Nuovo Cimento {\bf 31}, 1376 (1964).

\bibitem{holtwijk}
T.~Holtwijk, {\it Electron Pair Production by Photons}, Thesis, University of Groningen, Netherlands (1960).

\bibitem{asterix2} 
ASTERIX Collaboration, M.~Ziegler {\it et al.}, 
Phys. Lett. {\bf B 206}, 151 (1988).

\bibitem{achasov} 
M.~N.~Achasov {\it et al.}, 
{\it Proc. HADRON 97}, eds.  S-U.~Chung and H.~J.~Willutzki, AIP
conf. proc. 432 (1998) p.~26.

\bibitem{caso}
C.~Caso {\it et al.}, European Physical Journal {\bf C 3}, 1 (1998).

\bibitem{Atherton}
H.~W.~Atherton, C.~Bovet, P.~Coet, R.~Desalvo, N.~Doble, R.~Maleyran, 
E.~W.~Anderson, G.~von Dardel, K.~Kulka, M.~Boratav, J.~W.~Cronin, and
B.~D.~Milliken, Phys. Lett. {\bf B 158}, 81 (1985). 

\bibitem{shwe}
 H.~Shwe, F.~M.~Smith, and W.~H.~Barkas, Phys. Rev. {\bf 125}, 1024 (1962). 

\bibitem{stamer}
 P.~Stamer, S.~Taylor, E.~L.~Koller, T.~Huetter, J.~Grauman, and D. Pandoulas, Phys. Rev. {\bf 151}, 1108 (1966). 


\end{references}
\end{document}